\documentclass[11pt]{article}
\usepackage{amssymb, amsthm, amsmath}
\usepackage{graphicx}
\usepackage{bm}
\usepackage{setspace}
\usepackage{times}
\usepackage{epstopdf}
\usepackage[export]{adjustbox}

\usepackage{filecontents}
\usepackage{subfig}
\usepackage{mwe}
\usepackage{url}
\newlength{\tempheight}
\newlength{\tempwidth}

\newcommand{\rowname}[1]
{\rotatebox{90}{\makebox[\tempheight][c]{\textbf{#1}}}}

\newcommand{\columnname}[1]
{\makebox[\tempwidth][c]{\textbf{#1}}}

\begin{document}
\doublespacing
\begin{center}
  {\Large {\bf Analysis of Annual Cyclone Frequencies over Bay of Bengal: Effect of 2004 Indian Ocean Tsunami}}\\ \vspace{12pt}
  { {\bf Arnab Hazra}}\\ \vspace{12pt}
  { {\bf Interdisciplinary Statistical Research Unit, Indian Statistical Institute}}\\ \vspace{12pt}
\end{center}

\begin{abstract}
\noindent This paper discusses the time series trend and variability of the cyclone frequencies over Bay of Bengal, particularly in order to conclude if there is any significant difference in the pattern visible before and after the disastrous 2004 Indian ocean tsunami based on the observed annual cyclone frequency data obtained by India Meteorological Department over the years 1891-2015. Three different categories of cyclones- depression ($<34$ knots), cyclonic storm ($34-47$ knots) and severe cyclonic storm ($>47$ knots) have been analyzed separately using a non-homogeneous Poisson process approach. The estimated intensity functions of the Poisson processes along with their first two derivatives are discussed and all three categories show decreasing trend of the intensity functions after the tsunami. Using an exact change-point analysis, we show that the drops in mean intensity functions are significant for all three categories. As of author's knowledge, no study so far have discussed the relation between cyclones and tsunamis. Bay of Bengal is surrounded by one of the most densely populated areas of the world and any kind of significant change in tropical cyclone pattern has a large impact in various ways, for example, disaster management planning and our study is immensely important from that perspective.  
\end{abstract}

\section{Introduction}

Tropical cyclones (TC) and tsunamis are two very different kind of natural phenomena both originating from the oceans. Several necessary precursor environmental criteria for the formation of a TC have been discussed in \cite{Gray1968} and the six globally accepted criteria for cyclogenesis are- 1. sea surface temperature (SST) is more than $80^0$F or $26.5^0$C at least to a depth of 60 meters, 2. moist layers near the mid-troposphere, 3. an atmosphere that cools rapidly with height and unstable to moist convection, 4. $5^0$ latitude difference with the equator due to the requirement of a non-negligible amount of Coriolis force, 5. near-surface disturbance with sufficient vorticity and 6. lower values of vertical wind shear at the site of cyclogenesis. On the other hand, tsunami is caused by the friction of two slowly moving tectonic plates beneath the ocean floors that creates a large amount of seismic energy released in the form of an earthquake. Thus, it is highly unlikely that a tsunami is caused by a TC but the energy released in a tsunami can potentially affect some of the six criteria for cyclogenesis, particularly in case the tsunami is of extremely high magnitude. The geological records actually show a number of instances where large and abrupt climate changes have taken place (\cite{Alley2003}) including sudden shifts in the frequencies of TCs (\cite{Liu2001}).

\begin{figure}[h]
	\centering
	\includegraphics[height = 2.5in, width = 3.5in]{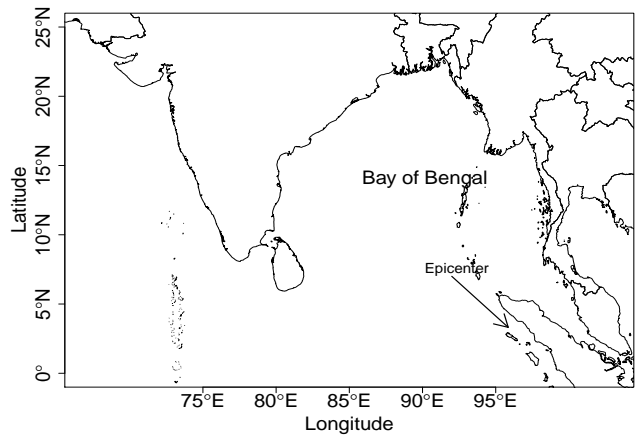}
	\caption{Stretch of the Bay of Bengal and the location of the epicenter of the 2004 Indian ocean tsunami.}
	\label{fig0}
\end{figure}

According to US Geological Survey report, the 2004 Indian ocean tsunami is the largest earthquake of past fifty years occurred on December 26, 2004 in which the India plate subducted beneath the Burma plate. It had a magnitude of 9.1 on the Richter scale with the epicenter located at (3.316$^0$N, 95.854$^0$E) near northwest of Sumatra and approximately 286,000 people were killed. The observed data from the Jason 1 altimetry satellite show an initial dominant wavelength of approximately 500 km and the shallow-water wave speed 750 km / hour approximately at 4.5 km depth (\cite{Gower2005}). The Bay of Bengal (BOB) is the northern extended arm of the Indian ocean on the east of India, stretched between latitudes 5$^0$N-2$2^0$N and longitudes 75$^0$E-100$^0$E and located very close to the epicenter (Figure \ref{fig0}) and hence it is likely that the tsunami would affect several water parameters in BOB. Considering the study period from November, 2004 to January, 2005, \cite{Reddy2009} concluded that the water temperature decreased by 2$^0$C-3$^0$C, salinity dropped by 2 psu along with specific changes in the pattern of surface geostrophic current just after the tsunami which became normal after a certain period.

Though the above references point out some short-period changes due to tsunami, long-term changes are not unlikely as well.  A method for detecting shifts in hurricane frequencies has been proposed by \cite{Elsner2004} 

Apart from just a change-point analysis, we also consider a proper a statistical modeling approach in order to model the annual cyclone frequencies. In the context of annual land-falling hurricane frequency analysis over the Atlantic coasts of United States, often along with the amount for the damages, several studies have considered HPP and non-homogeneous Poisson process (NHPP) along with some continuous stochastic models for the amount of economic losses (\cite{Katz2002}). 
In case the data associated with wind speeds, El Nino, the number of sunspots etc. are available, they are often used as covariates for modeling the mean structure (\cite{Katz2002}, \cite{Jagger2006}, \cite{Jagger2011}). While all these approaches assume a HPP model, \cite{Xiao2015} assumes a NHPP model for the seasonal point process approach, not from the perspective of a change-point analysis. \cite{Elsner2004} and \cite{Robbins2011} have used change-point detection techniques and conclude about the significant storm frequency increase during 1960--1995.

where a time $T^*$ is defined as a change-point if the counts arise from a homogeneous Poisson process (HPP) up to time $T^*$ and they arise from another HPP but with a shifted intensity parameter after $T^*$. In case we have long records available, e.g. records over 200 years and the 150th year is the suspected change-point, it is unlikely that the intensity function remains constant over first 150 years as different possible factors of cyclogenesis are likely to change, possibly drastically and hence the homogeneous assumption of the Poisson processes on both sides of $T^*$ is questionable.

As of author's knowledge, no such statistical literature is available which analyses the annual variability of frequencies of cyclones formed over BOB but a few studies are available where some preliminary analyses are done based on histograms, simple linear regression model etc. The annual frequencies of D over BOB is studied by \cite{Rao1958} but fails to notice any trend and the periodogram analysis also fails to notice any periodicity. 
\cite{Singh1999} studies the relation between the annual frequency of D and CS and the monsoon deficiency, \cite{Joseph1999} analyses the frequencies of D and CS using harmonic analysis and concludes that there is a very little long-term trend while it has an oscillation of period 36 years. Using separate linear regression models, \cite{Singh2000} compares the rates of changes of different categories of cyclones over time and notices that  the frequency of SCS increases at a faster rate compared to tropical depressions and also concludes about an oscillation period of 29 years.  \cite{Srivastava2000} again notices a decreasing trend in D over BOB during 1961-1998. The study of \cite{Singh2001} reveals that the frequencies of D and CS have decreased at a rate of 6-7 per hundred years and 1-2 per hundred years respectively. The recent study by \cite{Patwardhan2001} fits a five-degree orthogonal polynomial to annual overall (BOB + Arabian Sea) cyclone frequencies of the category CS and notices a significant downward trend. \cite{Niyas2009} notices that the decadal variability of CS and SCS are different and there is a significant long-term decreasing trend.

In this paper, we model the cyclone frequencies of different categories using non-homogeneous Poisson processes. The intensity functions along with their first two derivatives are estimated using a semi-parametric Poisson regression approach. The 95\% confidence intervals based on normal approximation are discussed. We are primarily interested in finding changes in the pattern of annual cyclone frequencies and hence study the trend and variability of the estimated functions before and after the tsunami. We notice drop in intensity for all three categories and also discussed in several studies. Finally, for testing the significance of the effect of tsunami, we conduct an exact a change-point analysis. As the intensity function $\lambda(t)$ is considered to be a non-constant function, considering the study-period to be $(a,b]$, we check if the overall mean intensity drops after the $T^*$, i.e. we test
\begin{eqnarray}
\noindent \nonumber && H_0: \frac{1}{T^* - a} \int_{a}^{T^*} \lambda(t) dt = \frac{1}{(b - T^*)} \int_{T^*}^{b} \lambda(t) dt~~ Vs~~ H_A: \frac{1}{T^* - a} \int_{a}^{T^*} \lambda(t) dt > \frac{1}{(b - T^*)} \int_{T^*}^{b} \lambda(t) dt.
\end{eqnarray}

In this paper, we concentrate on the application to time series dataset of annual frequencies of depressions (D), cyclonic storms (CS) and severe cyclonic storms (SCS)-- different categories of tropical cyclones defined by India Meteorological Department (IMD) forming over BOB. Land-falling cyclones lead to high death toll along with high economic damage over the coastline of India stretched over 7,516.6 Kilometers where 14.2\% of the total population (171 millions; Census 2014) of India reside (\url{http://iomenvis.nic.in}). Some of the recent deadliest cyclones are The Great 1970 Bhola Cyclone (in year 1970; the deadliest cyclone ever recorded; affected countries-- India and Bangladesh; 0.3--0.5 million people died from the storm and its after-effects and \$86.4 million property damage), ARB02 (in year 1998; more than 10,000 deaths; more than \$3 billion property damage), 1999 Odisha super cyclone (more than 10,000 deaths and \$4.5 billion property damage), YEMYIN (in year 2007; 983 deaths and \$2.1 billion property damage), HUDHUD (in year 2014; 124 deaths and \$3.4 billion property damage) etc. (\cite{Sommer1972}, The technical reports-- ``1998 Natural Catastrophes", ``Natural catastrophes and man-made disasters in 2008: North America and Asia suffer heavy losses" and from the NASA website for Hurricanes/Tropical Cyclones). 


Also, the Government of India carries on a budget plan in every five years and from the viewpoint of disaster management, the forecast cyclone frequencies for five future years would be very beneficial. 

The paper is organized as follows. In Section \ref{cyclone}, we discuss about the annual cyclone frequency dataset and necessity of a flexible non-homogeneous Poisson process approach. The statistical methods for estimation and hypothesis testing are discussed in Section \ref{stat_background}. In Section \ref{results}, we discuss the findings. Section \ref{conclusion} concludes.

\section{Cyclone data}
\label{cyclone}

\begin{figure}[h]
\centering
\includegraphics[height = 2in, width = 0.32\linewidth]{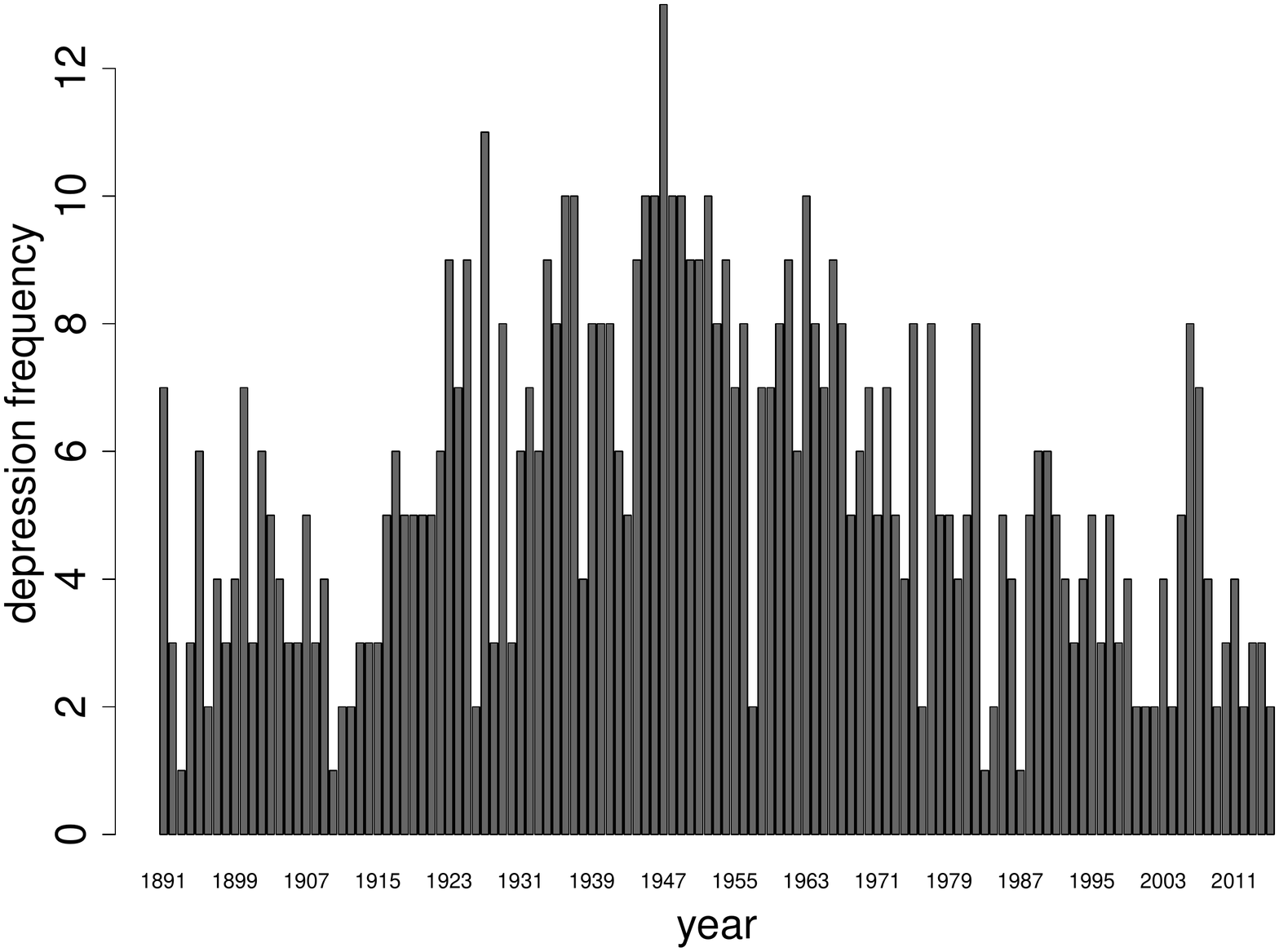}
\includegraphics[height = 2in, width = 0.32\linewidth]{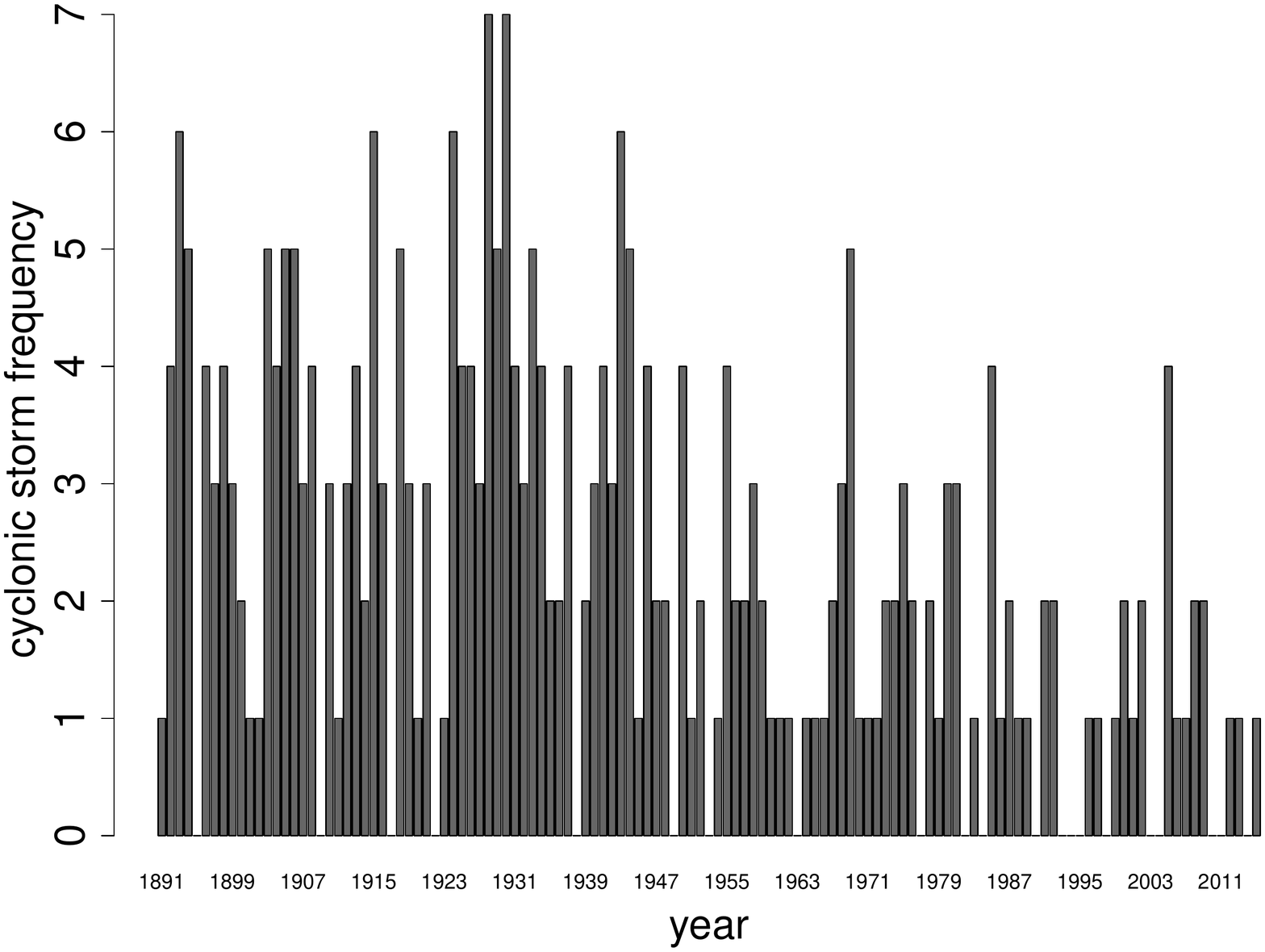}
\includegraphics[height = 2in, width = 0.32\linewidth]{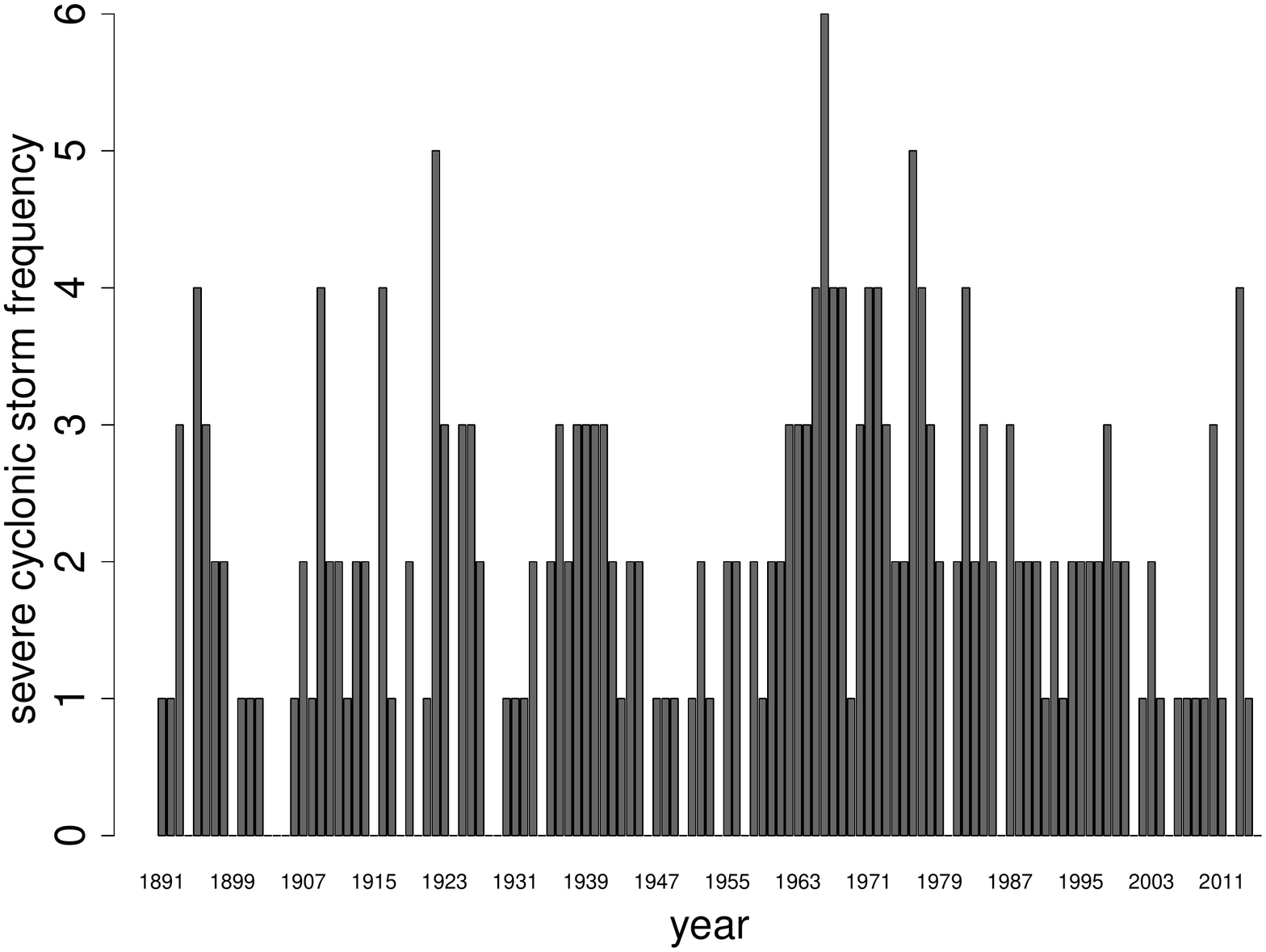}
\caption{Histograms of the annual frequencies of depressions (on the first panel), cyclonic storms (on the second panel) and severe cyclonic storms (on the third panel).}
\label{fig1}
\end{figure}

We analyze the annual frequencies of different categories of cyclones-- depressions (D; $<$34 knots; associated wind speed $<$60 km/hr), cyclonic storms (CS; 34-47 knots; associated wind speed 60-90 km/ hr) and severe cyclonic storms (SCS; $>$47 knots; associated wind speed $>$90 km/hr) forming over the Bay of Bengal (BOB) for the years 1891 to 2015. The data has been collected from the Cyclone e-Atlas published by IMD (\url{http://www.imd.gov.in}). Over the 125 years, total 682 D, 280 CS and 227 SCS have formed over BOB, AS and LAND respectively. More than 60\% cyclones forming over BOB reach Indian land area, around 30\% strike the coast of Bangladesh and Myanmar and the rest of 13\% dissipate over the sea (\url{http://www.rsmcnewdelhi.imd.gov.in}). The time series of annual frequencies of different categories of cyclones are provided in Figure \ref{fig1}. It is clearly visible that the frequencies highly change over the years and it is reasonable to assume that the rate parameter of the Poisson process to be non-constant and also the trend is highly non-linear requiring semi-parametric / non-parametric modeling of the intensity function. The values of average D, CS and SCS before and after the tsunami are provided in Table \ref{table1} and the drop in average frequencies are clearly visible, requiring a change-point analysis to conclude whether the drop is significant or not.

\begin{table}[ht]
\centering
\begin{tabular}{rrrr}
  \hline
Average frequency & D & CS & SCS \\ 
  \hline
Before tsunami & 5.6053 & 2.3421 & 1.8772 \\ 
  After tsunami & 3.9091 & 1.1818 & 1.1818 \\ 
   \hline
\end{tabular}
\caption{Average frequencies of depressions, cyclonic storms and severe cyclonic storms before and after the tsunami.}
\label{table1}
\end{table}

\section{Background statistical theory}
\label{stat_background}

We assume that the count of cyclone occurrences follow a non-homogeneous Poisson process model over the time interval $(a, b]$, i.e., we assume that no two cyclone occur at the same time and the time lag between two consecutive cyclones are independent and follow an exponential distribution with time-dependent rate parameter, say $\lambda(t)$ at time $t$. On the other hand, a homogeneous Poisson process assumes that the rate $\lambda(t)$ remains constant over time. In a simple explanation, $\lambda(t)$ represents the average frequency of cyclones at time $t$, e.g. saying 5.5 per year in 1985. The interval $(a,b]$ it can be written as a disjoint union of $N$ sub-intervals of equal length as $(a, b] = \cup_{n=1}^{N}(a + (n-1) \Delta, a + n \Delta]$ where $\Delta = (b-a)/N$ (e.g., in our study period of 125 years, $N=125$ and $\Delta$ represents a year) and the counting process $X_n$ denotes the number of events taking place within the interval $(a + (n-1) \Delta, a + n \Delta]$. In our set-up, $X_n$ denotes the cyclone frequency of the $n$-th year. Thus, $$X_n \overset{indep}{\sim} Poisson\left( \int_{a + (n-1) \Delta}^{a + n \Delta} \lambda(t) dt \right)$$ follows from the theory of the non-homogeneous Poisson process and also independent across $n = 1,2, \ldots, N$. Without assuming any parametric form for the rate function $\lambda(t)$, we assume it to be smooth, remain fairly constant within any interval $(a + (n-1) \Delta, a + n \Delta]$ so that $\log \left[\int_{a + (n-1) \Delta}^{a + n \Delta} \lambda(t) dt \right] \approx \int_{a + (n-1) \Delta}^{a + n \Delta} \log \left[\lambda(t)\right] dt$. Here an exact equality holds if and only if $\lambda(t)$ remains constant over the interval $(a + (n-1) \Delta, a + n \Delta]$ follows from Jensen's inequality. This assumption is quite unrealistic as the intensity are higher during the monsoon season compared to other months. But as we are interested in long-term changes and not in monthly changes, such a smoothness assumption does not affect our inference. Finally, we approximate the logarithm of the unknown function $\lambda(t)$ as a linear combination of cubic B-splines (\cite{DeBoor1978}), i.e. $\log \left[\lambda(t)\right] \approx \sum_{l=1}^L \beta_l B_l(t)$. Thus, $X_n \overset{indep}{\sim} Poisson(\mu_n)$ where $\log \left[\mu_n\right] \approx \sum_{l=1}^L \beta_l \int_{a + (n-1) \Delta}^{a + n \Delta} B_l(t) dt$ and hence the problem boils down to a Poisson regression set-up with covariates $\int_{a + (n-1) \Delta}^{a + n \Delta} B_l(t) dt; l=1, \ldots, L$ with natural logarithm as the link function and regression coefficients $\beta_l; l=1, \ldots, L$.

\subsection{Inference}
Based on the data, we estimate the vector $\beta = (\beta_1, \ldots, \beta_L)'$ using maximum likelihood estimates (MLEs). Suppose the estimated vector is given by $\hat{\beta} = (\hat{\beta_1}, \ldots, \hat{\beta_L})'$. Under some regularity conditions (\cite{Fahrmeir1985}), $\hat{\beta} \sim AN\left( \beta, I_{\beta}^{-1} \right)$ where $AN$ stands for asymptotic normal distribution and the information matrix $I_{\beta}$ is given by $$I_{\beta}^{ij} = \sum_{n=1}^{N} \int_{a + (n-1) \Delta}^{a + n \Delta} B_i(t) dt \times \int_{a + (n-1) \Delta}^{a + n \Delta} B_j(t) dt \times \exp \left[ \sum_{l=1}^L \beta_l \int_{a + (n-1) \Delta}^{a + n \Delta} B_l(t) dt \right]$$ where $I_{\beta}^{ij}$ denotes the $(i,j)$-th element of $I_{\beta}$. So,$\hat{\lambda}(t) = \exp \left[ \sum_{l=1}^L \hat{\beta}_l B_l(t) \right]$. Also we discuss about $\frac{\delta \lambda(t)}{\delta t}$ and $\frac{\delta^2 \lambda(t)}{\delta t^2}$, the first two derivatives of the intensity function $\lambda(t)$ given by

\begin{eqnarray}
\nonumber && \frac{\delta \lambda(t)}{\delta t} = \exp \left[ \sum_{l=1}^L \beta_l B_l(t) \right] \sum_{l=1}^L \beta_l \frac{\delta B_l(t)}{\delta t} \overset{Notation}{=} \lambda'(t) \\
\nonumber && \frac{\delta^2 \lambda(t)}{\delta t^2} = \exp \left[ \sum_{l=1}^L \beta_l B_l(t) \right] \left \lbrace \sum_{l=1}^L \beta_l \frac{\delta^2 B_l(t)}{\delta t^2} + \left(\sum_{l=1}^L \beta_l \frac{\delta B_l(t)}{\delta t} \right)^2 \right \rbrace \overset{Notation}{=} \lambda{''}(t)
\end{eqnarray}
and the estimated profiles $\hat{\lambda}'(t)$ and $\hat{\lambda}{''}(t)$ are obtained by replacing $\beta$ with $\hat{\beta}$.

For constructing the 95\% approximately normal distributed confidence intervals of the estimates of the functions $\lambda(t)$, $\lambda'(t)$ and $\lambda{''}(t)$, the variance estimates are obtained by Delta method (\cite{Cramer1946}) as follows
\begin{eqnarray}
\nonumber && Var\left[\hat{\lambda}(t) \right] \approx \frac{\delta \lambda(t)}{\delta \beta}^T I_{\beta}^{-1} \frac{\delta \lambda(t)}{\delta \beta}
; Var\left[\hat{\lambda}'(t) \right] \approx \frac{\delta \lambda'(t)}{\delta \beta}^T I_{\beta}^{-1} \frac{\delta \lambda'(t)}{\delta \beta}; Var\left[\hat{\lambda}^{''}(t) \right] \approx \frac{\delta \lambda^{''}(t)}{\delta \beta}^T I_{\beta}^{-1} \frac{\delta \lambda^{''}(t)}{\delta \beta}
\end{eqnarray}
where $\frac{\delta \lambda(t)}{\delta \beta} = \left( \frac{\delta \lambda(t)}{\delta \beta_1}, \ldots, \frac{\delta \lambda(t)}{\delta \beta_L} \right)^T$, $\frac{\delta \lambda'(t)}{\delta \beta} = \left( \frac{\delta \lambda'(t)}{\delta \beta_1}, \ldots, \frac{\delta \lambda'(t)}{\delta \beta_L} \right)^T$ and $\frac{\delta \lambda^{''}(t)}{\delta \beta} = \left( \frac{\delta \lambda^{''}(t)}{\delta \beta_1}, \ldots, \frac{\delta \lambda^{''}(t)}{\delta \beta_L} \right)^T$. The elements of the vectors are calculated as follows
\begin{eqnarray}
\nonumber \frac{\delta \lambda(t)}{\delta \beta_l} &=& \exp \left[ \sum_{l=1}^L \beta_l B_l(t) \right] B_l(t) \\
\nonumber \frac{\delta \lambda'(t)}{\delta \beta_l} &=& \exp \left[ \sum_{l=1}^L \beta_l B_l(t) \right] \left \lbrace \frac{\delta B_l(t)}{\delta t} + B_l(t) \sum_{l=1}^L \beta_l \frac{\delta B_l(t)}{\delta t} \right \rbrace \\
\nonumber \frac{\delta \lambda{''}(t)}{\beta_l} &=& \exp \left[ \sum_{l=1}^L \beta_l B_l(t) \right] \left \lbrace \frac{\delta^2 B_l(t)}{\delta t^2} \right. \\
\nonumber && \left. + B_l(t) \left[\sum_{l=1}^L \beta_l \frac{\delta^2 B_l(t)}{\delta t^2} + \left(\sum_{l=1}^L \beta_l \frac{\delta B_l(t)}{\delta t} \right)^2 \right] + 2 \frac{\delta B_l(t)}{\delta t} \sum_{l=1}^L \beta_l \frac{\delta B_l(t)}{\delta t} \right \rbrace.
\end{eqnarray}

\subsection{Hypothesis testing}

In case data are available at an annual grid, it is not possible to detect a change-point uniquely within an year. Suppose there is a change-point suspected at $a+ K\Delta$, i.e., at the end of the $K$-th year. Thus, we divide the domain into two parts- $(a, a+ K\Delta] = \cup_{n=1}^{K}(a + (n-1) \Delta, a + n \Delta]$ and $(a+ K\Delta, b] = \cup_{n=K+1}^{N}(a + (n-1) \Delta, a + n \Delta]$. We are interested in testing whether the overall mean of the intensity function $\lambda(t); t \in (a, b]$ remains same before and after the change point $a+ K\Delta$ or whether the average intensity drops after the change point, i.e. mathematically
\begin{eqnarray}
\nonumber && H_0: \frac{1}{K \Delta} \int_{a}^{a + K \Delta} \lambda(t) dt = \frac{1}{(N-K) \Delta} \int_{a + K \Delta}^{b} \lambda(t) dt \\
\nonumber && Versus~~~~ H_A: \frac{1}{K \Delta} \int_{a}^{a + K \Delta} \lambda(t) dt > \frac{1}{(N-K) \Delta} \int_{a + K \Delta}^{b} \lambda(t) dt.
\end{eqnarray}

\noindent Now, $\sum_{n=1}^{K} X_n \sim Poisson\left(\sum_{n=1}^K \int_{a + (n-1) \Delta}^{a + n \Delta} \lambda(t) dt \right)$ and $\sum_{n=K+1}^{N} X_{n} \sim Poisson\left(\sum_{n=K+1}^N \int_{a + (n-1) \Delta}^{a + n \Delta} \lambda(t) dt \right)$. So, the conditional distribution of $\sum_{n=1}^K X_n$ given $\sum_{n=1}^N X_n$ is given by $$\sum_{n=1}^K X_n | \sum_{n=1}^N X_n \sim Binomial \left(\sum_{n=1}^N X_n, \frac{\int_{a}^{a + K \Delta} \lambda(t) dt}{\int_{a}^{b} \lambda(t) dt} \right).$$ Under $H_0$, $\sum_{n=1}^K X_n | \sum_{n=1}^N X_n \sim Binomial \left(\sum_{n=1}^N X_n, \frac{K}{N} \right)$ and we reject $H_0$ if $\sum_{n=1}^K X_n$ is large. The $p$-value is calculated as $p = 1 - p_B\left(\sum_{n=1}^K X_n,  \sum_{n=1}^N X_n, \frac{K}{N}\right)$ where $p_B$ denotes the binomial distribution function with size $\sum_{n=1}^N X_n$ and success probability $ \frac{K}{N}$ evaluated at $\sum_{n=1}^K X_n$. Hence for a test of level 0.05, we reject $H_0$ if $p$ is less than 0.05.

\section{Results and Discussions}
\label{results}

\begin{figure}[h]
\centering
\includegraphics[height = 2in, width = 0.32\linewidth]{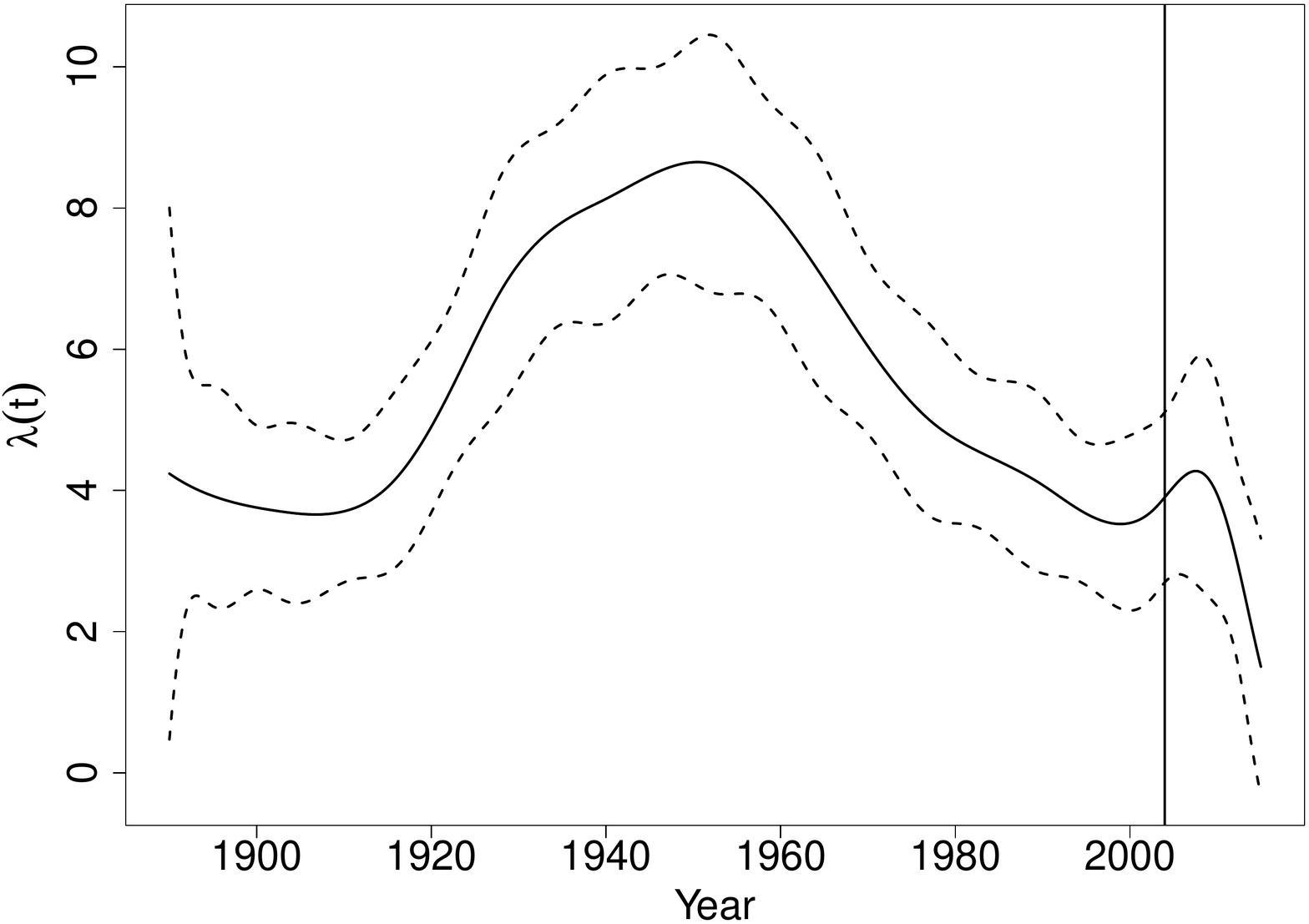}
\includegraphics[height = 2in, width = 0.32\linewidth]{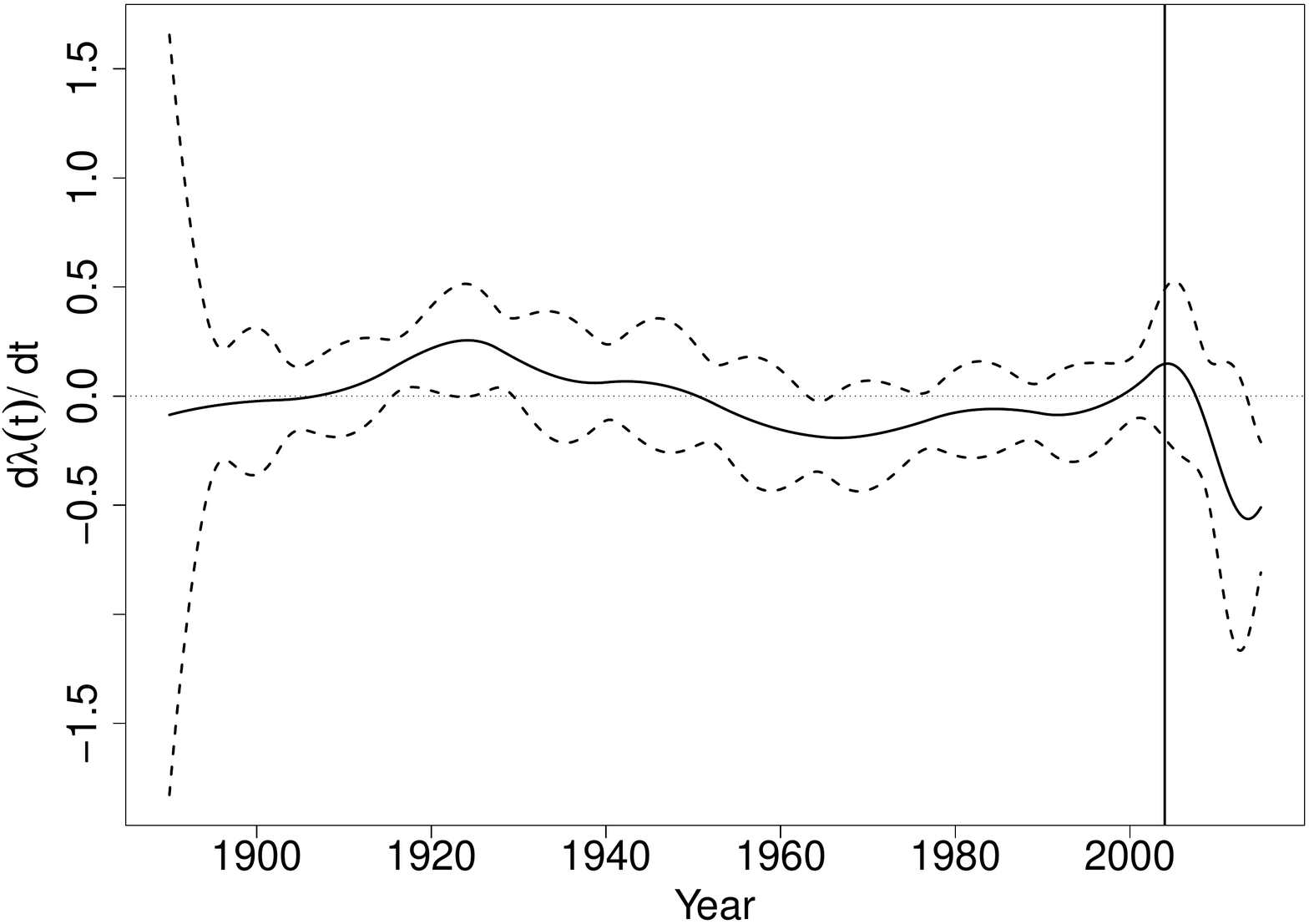}
\includegraphics[height = 2in, width = 0.32\linewidth]{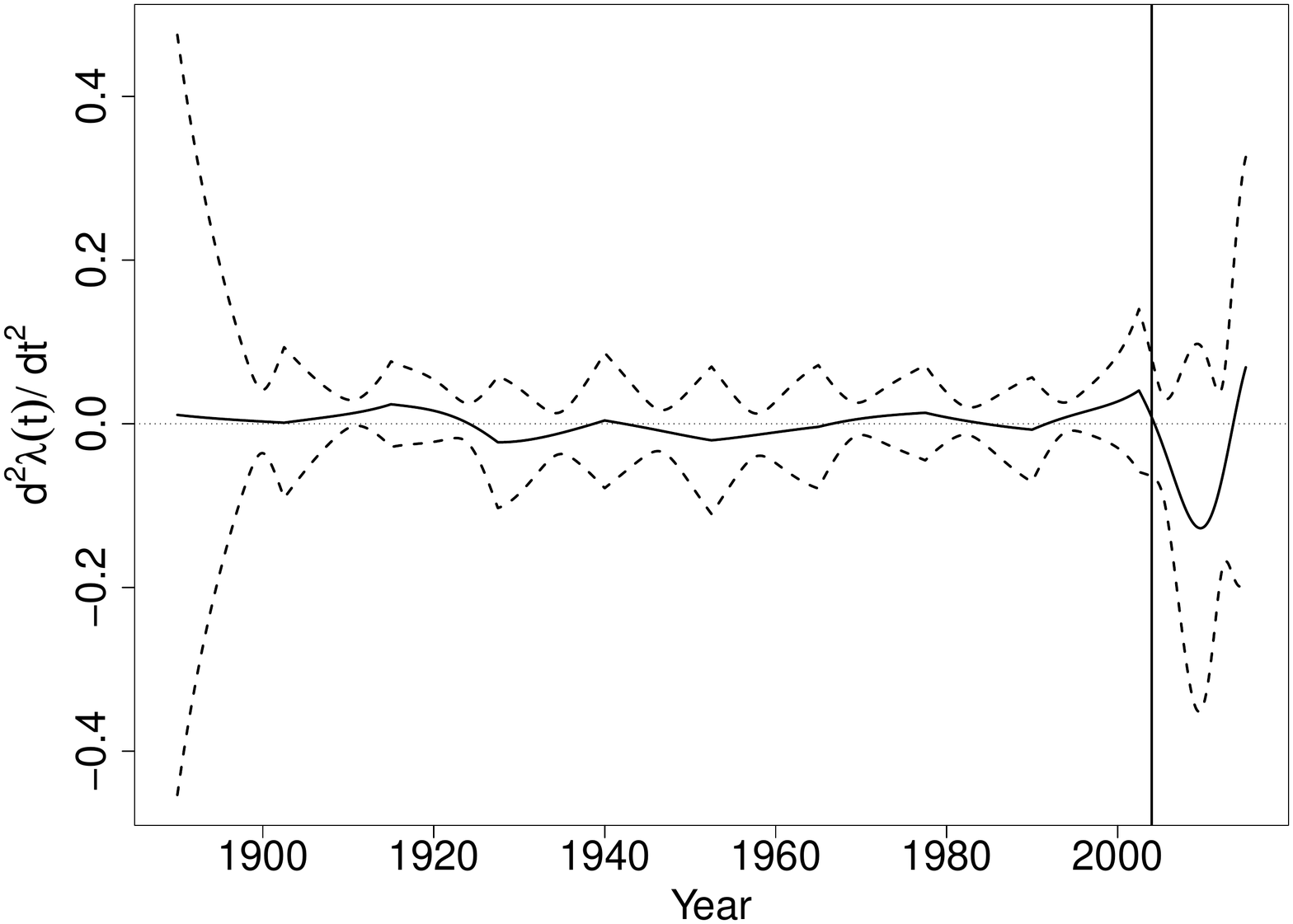}
\caption{Estimates of the functions $\lambda(t)$, $\lambda'(t)$ and $\lambda{''}(t)$ along with 95\% confidence intervals corresponding to the frequencies of depressions. The vertical line shows the time of tsunami.}
\label{fig2}
\end{figure}

\begin{figure}[h]
\centering
\includegraphics[height = 2in, width = 0.32\linewidth]{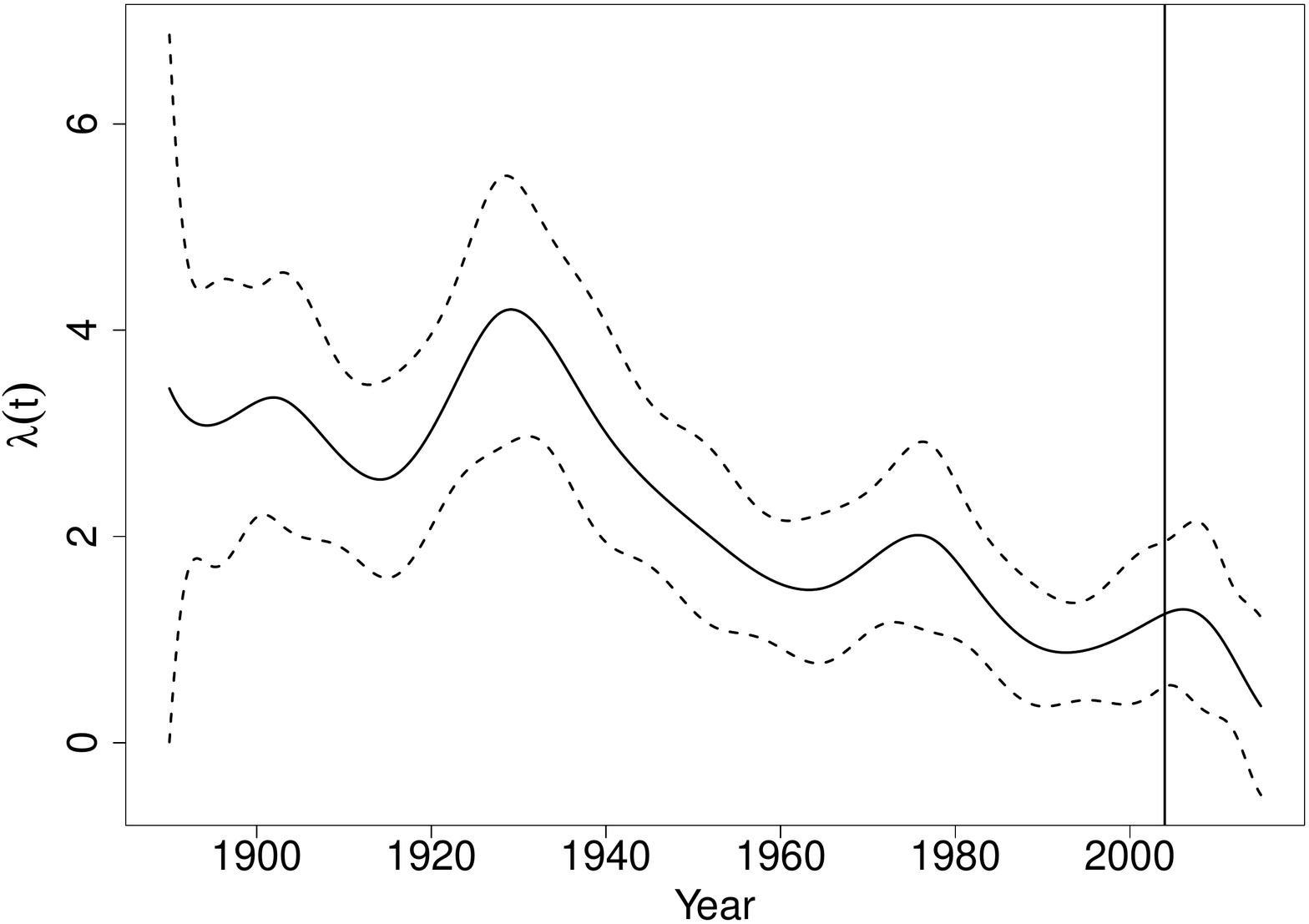}
\includegraphics[height = 2in, width = 0.32\linewidth]{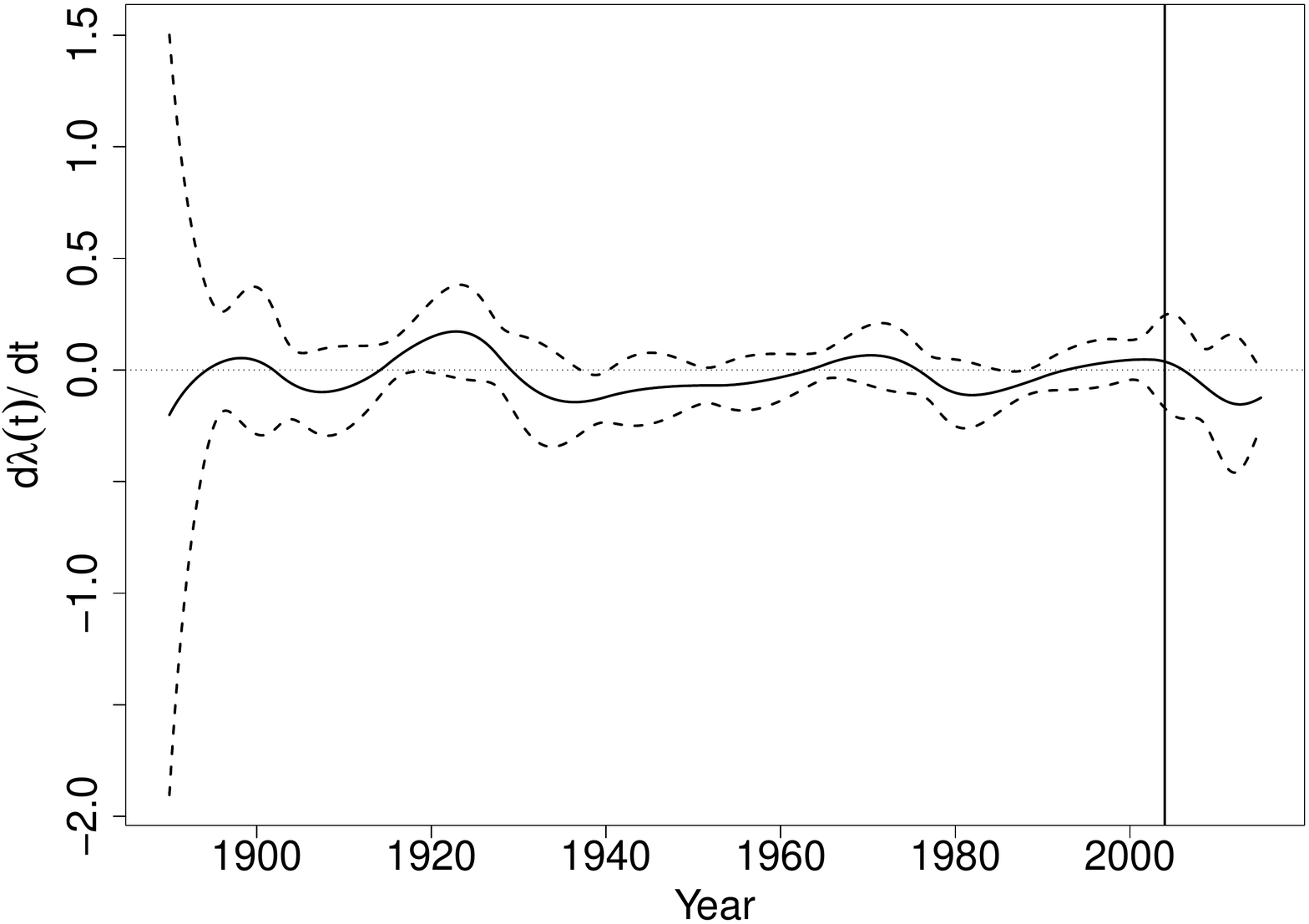}
\includegraphics[height = 2in, width = 0.32\linewidth]{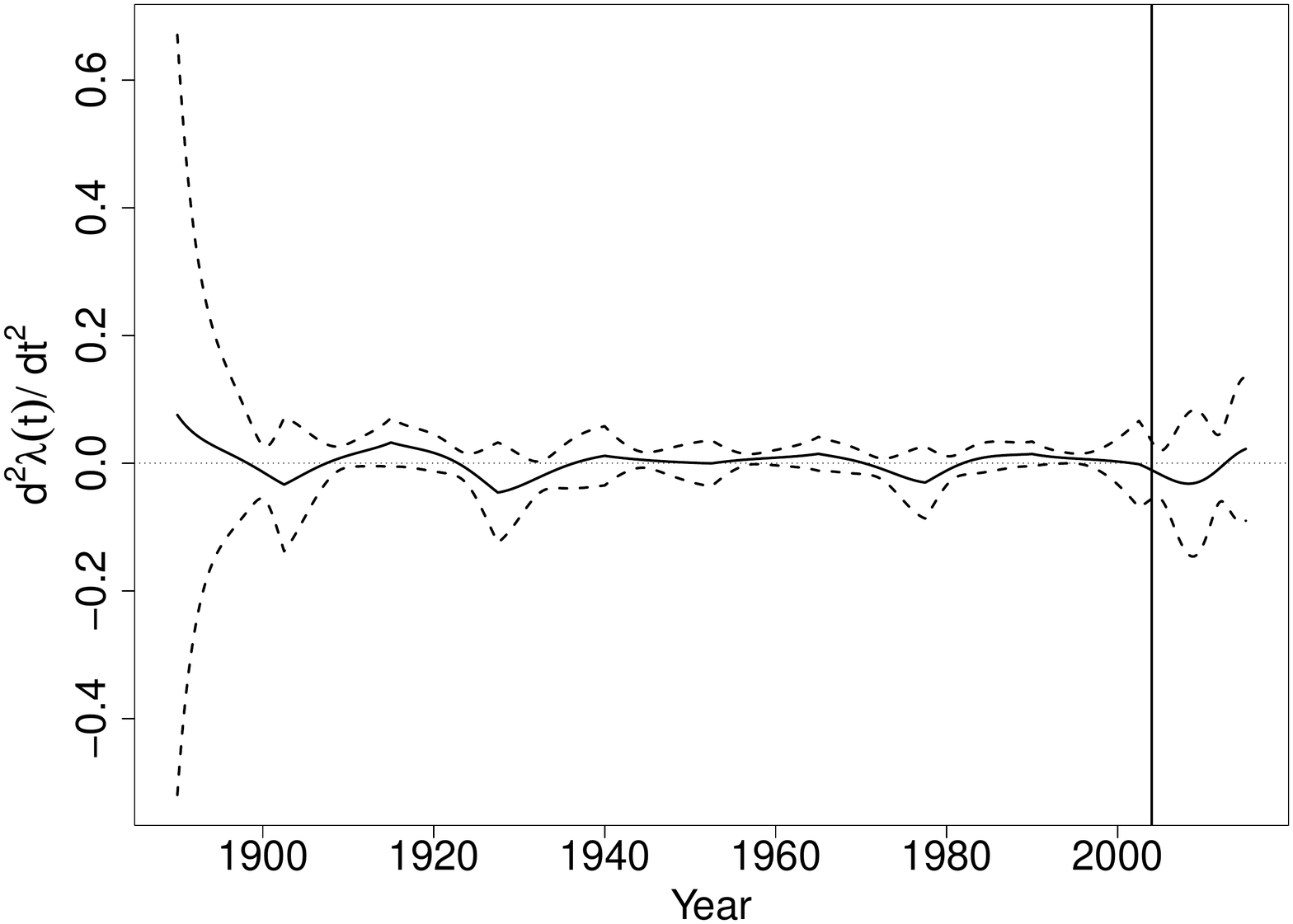}
\caption{Estimates of the functions $\lambda(t)$, $\lambda'(t)$ and $\lambda{''}(t)$ along with 95\% confidence intervals corresponding to the frequencies of cyclonic storms. The vertical line shows the time of tsunami.}
\label{fig3}
\end{figure}

\begin{figure}[h]
\centering
\includegraphics[height = 2in, width = 0.32\linewidth]{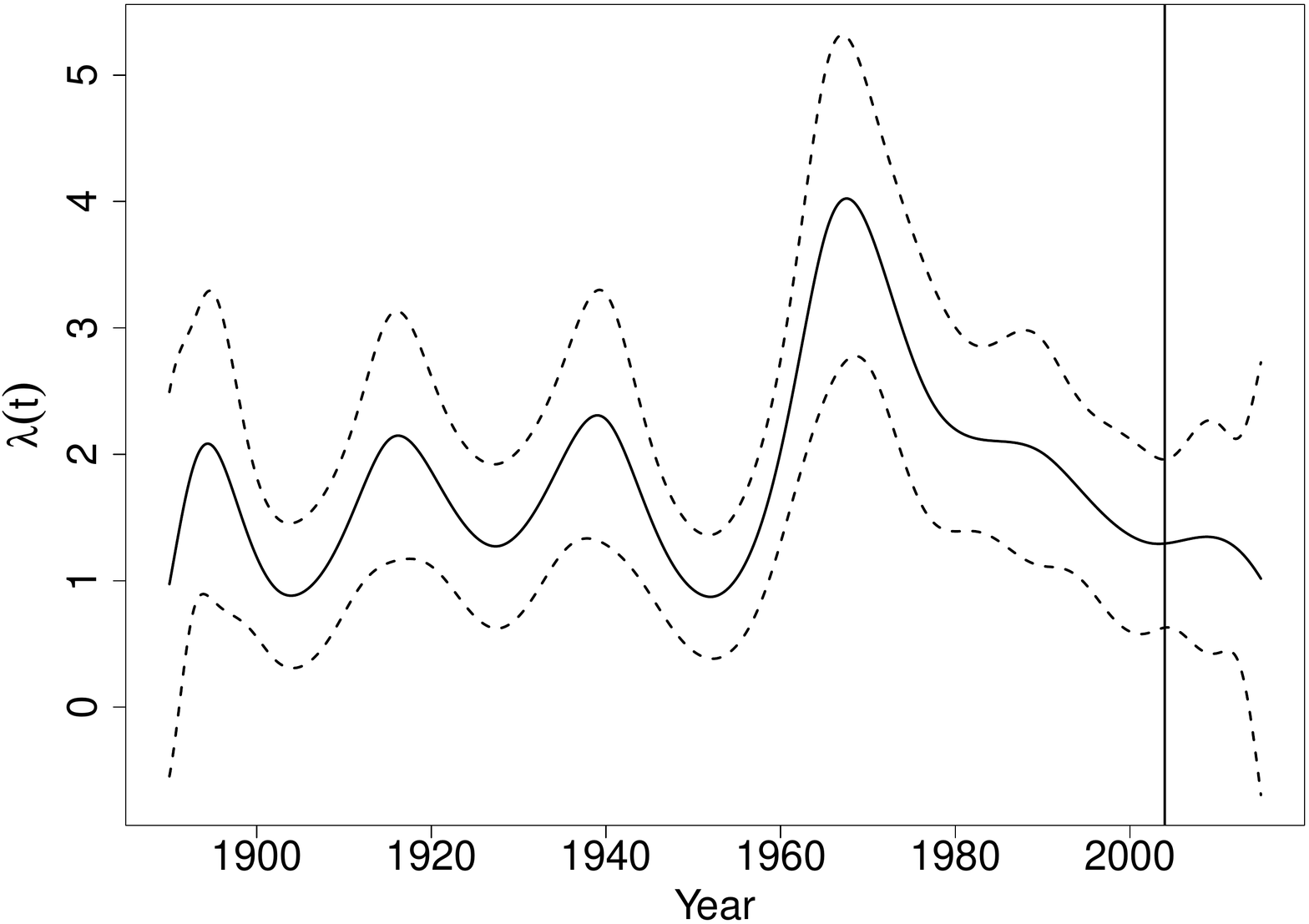}
\includegraphics[height = 2in, width = 0.32\linewidth]{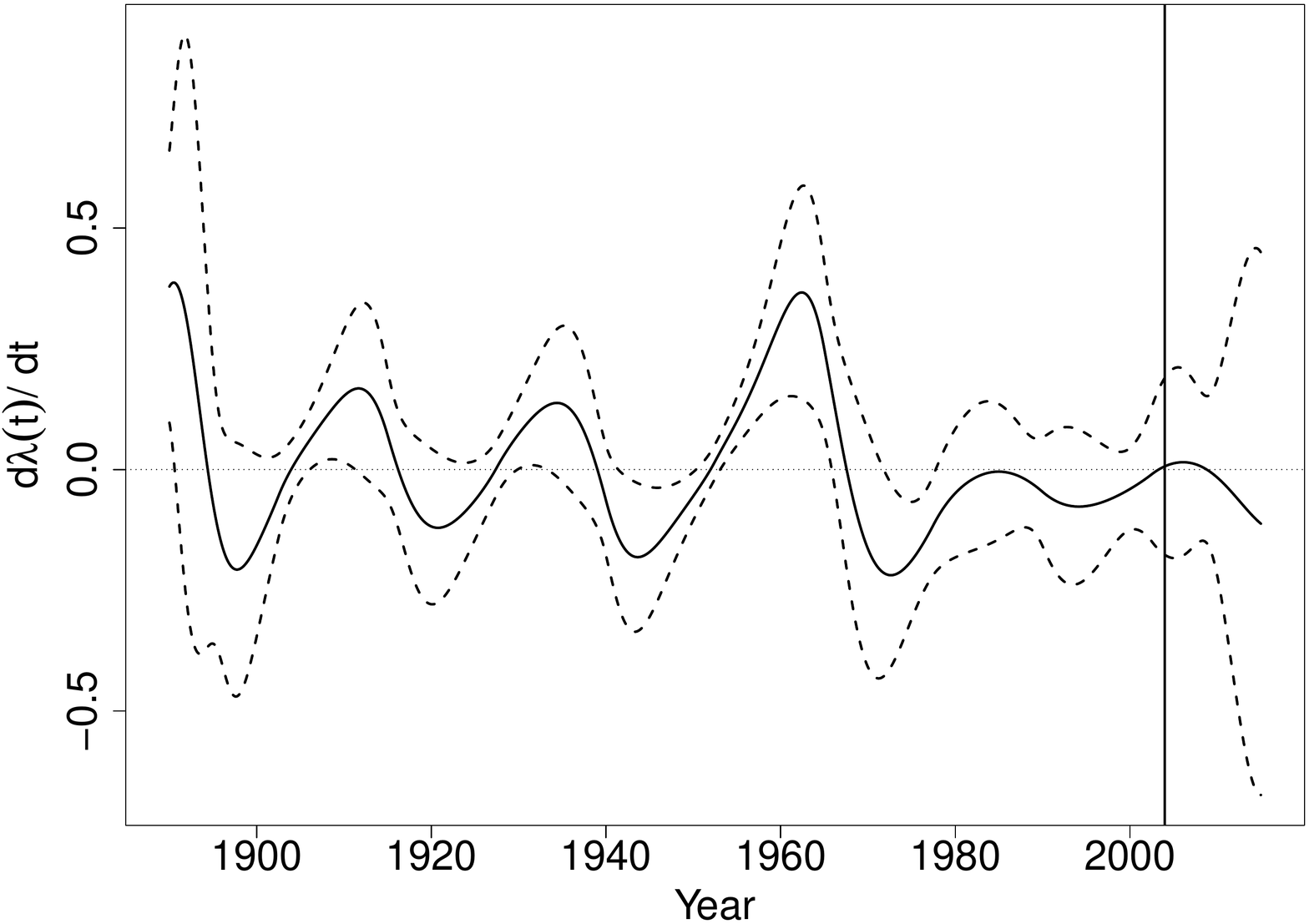}
\includegraphics[height = 2in, width = 0.32\linewidth]{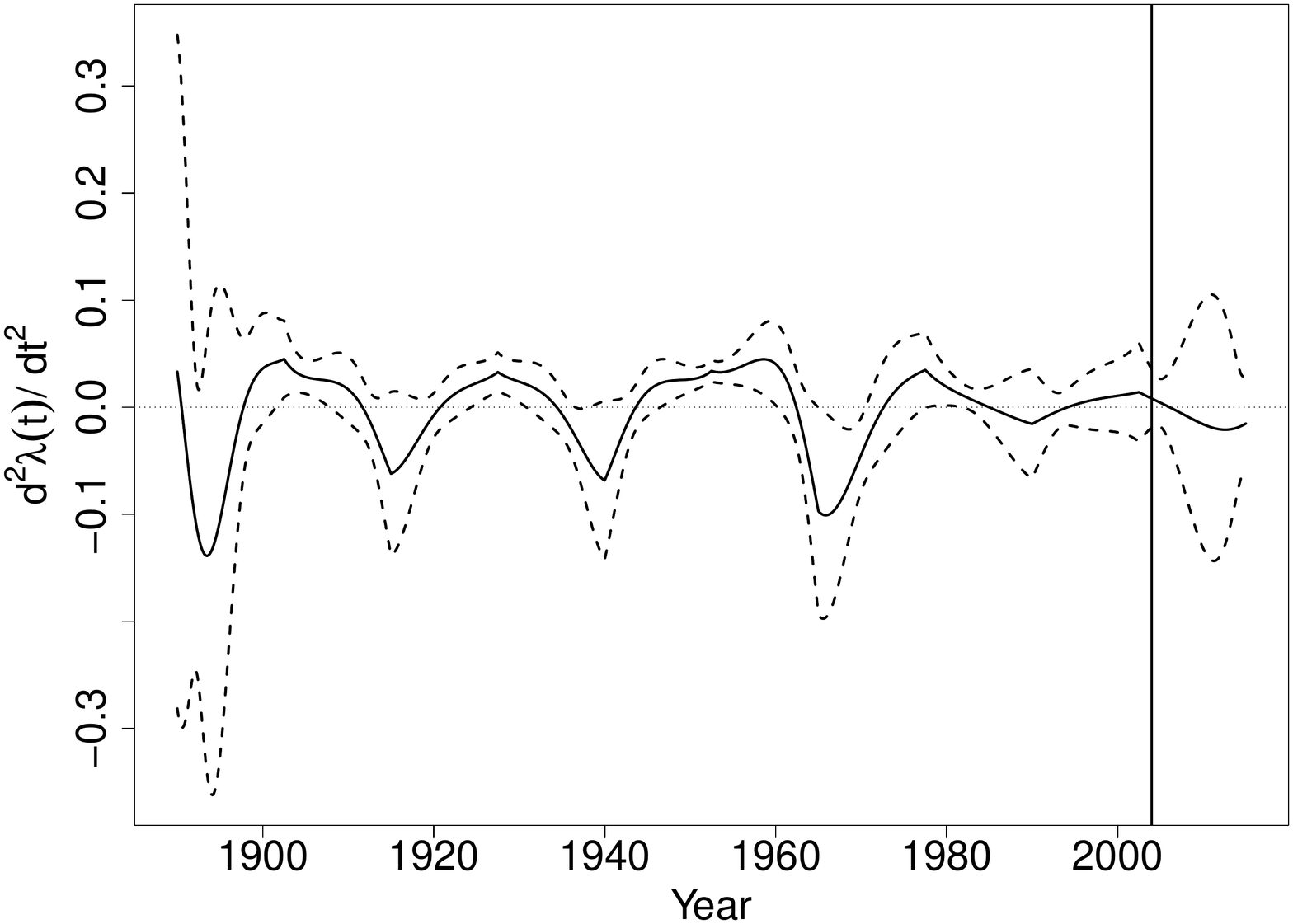}
\caption{Estimates of the functions $\lambda(t)$, $\lambda'(t)$ and $\lambda{''}(t)$ along with 95\% confidence intervals corresponding to the frequencies of severe cyclonic storms. The vertical line shows the time of tsunami.}
\label{fig4}
\end{figure}

The estimates of the functions $\lambda(t)$, $\lambda'(t)$ and $\lambda{''}(t)$ along with 95\% confidence intervals based on normal approximation for the cases D, CS and SCS are provided in Figures \ref{fig2}, \ref{fig3} and \ref{fig4} respectively. From the left panel of Figure \ref{fig2}, it is clearly visible that the rate increases within the period 1910-1950 and then decreases within the period 1950-2000 and then after increasing for a few years, it drops rapidly at the end. From the second panel, it is clear that the derivative $\lambda'(t)$ remains positive within 1910-1950 and negative within 1950-2000 conforming with the left panel and drops rapidly after the end of 2004, exactly the time when the tsunami occurred. From the right panel, it is clear that the pattern in $\lambda{''}(t)$ is highly non-conformable before and after the incident. The slight difference between the time of tsunami and when the sudden drop in $\lambda{''}(t)$ starts is because of the approximation by B-spline functions. The inflated shape of the confidence intervals between two knots of the B-spline functions in the middle and right panels of Figure \ref{fig2} is due to the larger variance of estimating the function at a point far from a knot and hence reasonable. As we model using cubic B-splines, their first two derivatives are quadratic and linear respectively and hence the difference in shape of the inflation in the two panels.

From the left panel of Figure \ref{fig3}, it is clear that the rate $\lambda(t)$ varies slightly within the period 1891-1910 with a overall dropping pattern, the increases within the period 1920-1930 and then decreases almost linearly. Notice that, the increasing trend in the decade before and after the tsunami are completely opposite.

Based on the proposed hypothesis testing procedure, we test the significance of change in the overall mean intensity function after the tsunami separately for each of the three categories of cyclone-D, CS and SCS and the p-values we have obtained are \textbf{0.0068}, \textbf{0.0028} and \textbf{0.0332} respectively; all values are below the level of the test, i.e. 0.05. Hence, we conclude that the change in the overall mean intensity function is significant for all three categories of the cyclone.

\section{Discussions and conclusions}
\label{conclusion}

We have analyzed the annual frequencies of three different categories of cyclones- depressions, cyclonic storms and severe cyclonic storms forming over Bay of Bengal over the years 1891-2015 using non-homogeneous Poisson processes. The time series plots of the cyclone frequencies show highly non-linear pattern across the years for all three categories. Hence, instead of assuming some parametric form of the intensity functions, we model them semi-parametrically by approximating the logarithm of the intensity function as a linear combination of cubic B-splines. This brings down the problem of estimating the intensity function in a Poisson regression set-up with natural logarithm link function, a very common tool in applied statistics. The B-spline coefficients are then estimated using maximum likelihood approach and further used to estimate the intensity functions along with their first two derivatives. Using delta method, the variances of the estimated intensity functions and their derivatives are calculated and consequently 95\% confidence intervals are obtained based on normal approximation. A less restricted definition of change-point, compared to the one given by Elsner et al. (2004), has been proposed for non-homogeneous Poisson processes and also an exact test is proposed for testing the significance of the change-point. The trends and variabilities in the estimated functions are discussed. In almost
all the cases, the estimates of $\lambda(t)$, $\lambda'(t)$ and $\lambda^{''}(t)$ show decreasing trend just after the tsunami. Based on the p-values obtained from the hypothesis testing, we conclude that the change-point due to tsunami is significant. 

By further analyzing the effect of tsunami on several factors of cyclogenesis, identifying the ones got affected significantly would be a future endeavor. A few studies are available (Reddy et al. , 2009; Ramlan et al. , 2014) which discuss the impact of the 2004 Indian ocean tsunami on water temperature, salinity etc. but they are focused on short-period changes. As of author’s knowledge, no article is available discussing the long-term impact on climate. 

Not just the 2004 Indian ocean tsunami but also from a large number of deadliest cyclones in Bay of Bengal, India and its neighboring countries- Bangladesh and Myanmar have faced high death tolls along with high economic damages. Some of the deadliest cyclones are The Great 1970 Bhola Cyclone (in year 1970; the deadliest cyclone ever recorded; affected countries– India and Bangladesh; 0.3–0.5 million people died from the storm and its after-effects and \$86.4 million property damage), 1999 Odisha super cyclone (more than 10,000 deaths and \$4.5 billion property damage), YEMYIN (in year 2007; 983 deaths and \$2.1 billion property damage), HUDHUD (in year 2014; 124 deaths and \$3.4 billion property damage) etc. (Sommer and Mosley (1972), The technical reports– “1998 Natural Catastrophes”, “Natural catastrophes and man-made disasters in 2008: North America and Asia suffer heavy losses” and from the NASA website for Hurricanes/Tropical Cyclones). Thus, from the viewpoint of a proper disaster management planning of India and its neighboring countries, a proper analysis of cyclone frequencies and their relation with a recent megathrust like 2004 Indian ocean tsunami would be very beneficial.


\begin{thebibliography}{9}

\bibitem{Niyas2009}
\textsc{Niyas, N.T.}, \textsc{A. K. Srivastava} and \textsc{Hatwar, H.R.} (2009). \textit{Variability and Trend in the Cyclonic Storms Over North Indian Ocean}, National Climate Centre, Office of the Additional Director General of Meteorology (Research), India Meteorological Department.

\bibitem{Gray1968}
\textsc{Gray, W.M.}  (1968). Global View of the Origin of Tropical Disturbances and Storms. \textit{Monthly Weather Review}  \textbf{68} 669--700.

\bibitem{Velasco1987}
\textsc{Velasco, I.} and \textsc{Fritsch, J.M.}  (1987). Mesoscale Convective Complexes in the Americas. \textit{Journal of Geophysical Research}  \textbf{92} 9591--9613.

\bibitem{Juarez2010}
\textsc{Juarez, M.L.A.}, \textsc{Mora, R.D.}, \textsc{Mariles, G.F.} and \textsc{Gutierrez-Lopez, A.} (2010).  Synthetic generation of monthly sea surface temperatures in “El Ni$\tilde{n}$o” regions by means of the Fiering-Svanidze method. \textit{Atmosfera} \textbf{23} 367--386.

\bibitem{Katz2002}
\textsc{Katz, R.W.} (2002). Stochastic modeling of hurricane damage. \textit{Journal of Applied Meteorology} \textbf{41} 754--762.

\bibitem{Jagger2006}
\textsc{Jagger, T.H.} and \textsc{Elsner, J.B.} (2006). Climatology models for extreme hurricane winds near the United States. \textit{Journal of Climate} \textbf{19} 3220--3236.

\bibitem{Xiao2015}
\textsc{Xiao, S.}, \textsc{Kottas, A} and \textsc{Sanso, B.} (2015) Modeling for seasonal marked point processes: An analysis of evolving hurricane occurrences. \textit{Ann. Appl. Stat.} \textbf{9} 353--382.

\bibitem{Jagger2011}
\textsc{Jagger, T.H.}, \textsc{Elsner, J.} and \textsc{Burch, R.} (2011). Climate and solar signals in property damage losses from hurricanes affecting the United States. \textit{Natural Hazards} \textbf{58} 541--557.

\bibitem{Elsner2004}
\textsc{Elsner, J.B.}, \textsc{Xu, F.N.} and \textsc{Jagger, T.H.} (2004). Detecting shifts in hurricane rates using a Markov Chain Monte Carlo approach. \textit{Journal of Climate} \textbf{17} 2652--2666. 

\bibitem{Rao1958}
\textsc{Rao, K.N.} and \textsc{Jayaraman, S.} (1958). A statistical study of frequency of depressions/cyclones in the Bay of Bengal. \textit{Indian Journal of Meteorology and Geophysics} \textbf{9} 233--250.

\bibitem{Robbins2011}
\textsc{Robbins, M.W.}, \textsc{Lund, R.B.}, \textsc{Gallagher, C.M.} and \textsc{Lu, Q.} (2011). Change-points in the North Atlantic tropical cyclone record. \textit{J. Amer. Statist. Assoc.} \textbf{106} 89--99.

\bibitem{Shyamala1996}
\textsc{Shyamala, B.} and \textsc{Iyer, B.G.} (1996). Statistical study of cyclonic disturbances in Arabian Sea. \textit{Proceedings to TROPMET-1996}, Vishakhapatnam, India.

\bibitem{Singh1999}
\textsc{Singh, O.P.} and \textsc{Rout, R.K.} (1999). Frequency of cyclonic disturbances over the North Indian Ocean during ENSO years. \textit{Proceedings of TROPMET-1999 Symposium} 297.

\bibitem{Joseph1999}
\textsc{Joseph, P.V.} and \textsc{Xavier, P.K.} (1999) Monsoon rainfall and frequencies of monsoon depressions and tropical cyclones of recent 100 years and an outlook for the first decades of the 21st century. \textit{Proceedings of TROPMET-1999 Symposium} 364.

\bibitem{Singh2000}
\textsc{Singh, O.P.}, \textsc{Khan, T.M.A.} and \textsc{Rahman, M.S.} (2000). Changes in the frequencies of Tropical cyclones over the North Indian Ocean. \textit{Meteorology and Atmospheric Physics} \textbf{75} 11--20.

\bibitem{Srivastava2000}
\textsc{Srivastava, A.K.}, \textsc{Sinha Ray, K.C.} and \textsc{De, U.S.} (2000) Trends in the frequency of cyclonic disturbances and their intensification over Indian seas. \textit{Mausam} \textbf{51} 113--118.

\bibitem{Singh2001}
\textsc{Singh, O.P.} (2001). Long term trends in the frequency of monsoonal cyclonic disturbances over the North Indian Ocean. \textit{Mausam} \textbf{52} 655--658.

\bibitem{Patwardhan2001}
\textsc{Patwardhan, S.K.} and \textsc{Bhalme, H.N.} (2001). A Study of Cyclonic Disturbances over India and the Adjacent Ocean. \textit{Int. J. Climatol.} \textbf{21} 527--534.

\bibitem{DeBoor1978}
\textsc{De Boor, C.}(1978). A practical guide to splines. \textit{Springer-Verlag New York} \textbf{27}.

\bibitem{Fahrmeir1985}
\textsc{Fahrmeir, L.} and \textsc{Kaufmann, H.} (1985) Consistency and asymptotic normality of the maximum likelihood estimator in generalized linear models. \textit{The Annals of Statistics} \textbf{13} 342-368.

@article{cramer1946methods,
	title={Methods of mathematical statistics},
	author={Cram{\'e}r, Harald},
	journal={Princeton: Princeton University Press. CramerMethods of Mathematical Statistics1946},
	year={1946}
}

\bibitem{Cramer1946}
\textsc{Cram{\'e}r, H.} (1946) Methods of mathematical statistics. \textit{Princeton University Press}.


\bibitem{Alley2003}
\textsc{Alley, R. B.}, \textsc{Alley, R. B.}, \textsc{Alley, R. B.}, \textsc{Alley, R. B.}, and \textsc{Kaufmann, H.} (1985) Abrupt climate change. \textit{The Annals of Statistics} \textbf{13} 342-368.



\bibitem{Liu2001}
Liu, K. B., Shen, C., \& Louie, K. S. (2001). A 1,000‐Year History of Typhoon Landfalls in Guangdong, Southern China, Reconstructed from Chinese Historical Documentary Records. Annals of the Association of American Geographers, 91(3), 453-464.

\bibitem{Gower2005}
Gower, J. (2005). Jason 1 detects the 26 December 2004 tsunami. Eos, Transactions American Geophysical Union, 86(4), 37-38.

\bibitem{Reddy2009}
Reddy, N., Aung, T. H., \& Singh, A. M. (2009). Effect of the 2004 ‘Boxing Day’Tsunami on water properties and currents in the Bay of Bengal. American Journal of Environmental Sciences, 5(3), 247-255.
\end{thebibliography}
\end{document}